\documentclass[12pt]{article}
\usepackage{latexsym}
\usepackage{amsfonts}
\usepackage{amssymb}
\usepackage[cp1250]{inputenc}

\title{Canonical formalism and quantization of perturbative sector of higher-derivative theories }
\author{K.Andrzejewski\thanks{supported by the grant 691 of University of Lodz and 
by The European Social Fund and Budget of State
(implemented) under The Integrated Regional Operational Programme }  ,
 K.Bolonek\thanks{supported by the grant  1 P03B 125 29 of the Polish Ministry of Science 
and  by The European Social Fund and Budget of State
(implemented) under The Integrated Regional Operational Programme.
},
 J.Gonera\thanks{supported by the grants 690 and 795 of University of Lodz}, P.Ma\'slanka \thanks{supported by the grant 1 P03B 02 128 of the Polish Ministry of Science.} \\ 
Department of Theoretical Physics II, Institute of Physics,\\
University of {\L}\'od\'z ,
Pomorska 149/153, 90 - 236 {\L}\'od\'z, Poland.}

\date{}

\begin{document}
\maketitle
\begin{abstract}
The theories defined by Lagrangians containing second time derivative are considered. It is shown that if the second derivatives enter 
only the terms multiplied by coupling constant one can consistently define the perturbative sector via Dirac procedure. The possibility 
of introducing standard canonical variables is analysed in detail. The ambiguities in quantization procedure are pointed out. 
\end{abstract}

\newpage
\section{Introduction}
Higher-derivative theories were introduced quite early in an attempt to regularize the ultraviolet divergencies of quantum field theories \cite{b1}.

Another contex in which higher-derivative and nonlocal theories appear naturally is the description of the low-energy phenomena in terms of 
effective action which is nonlocal as a result of integrating out high energy degrees of freedom \cite{b2}. Moreover, theories with infinite 
degree derivatives do appear in the framework of string theory \cite{b3}, \cite{b4} and as a modified theories of gravity \cite{b5}.

In recent years the emergence of noncommutative field theories \cite{b6} has revived the discussion concerning higher-derivative theories. 
Apart from their string-theoretical origin noncommutative field theories can be viewed as an attempt to describe the dynamics at the scales 
where the very notion of space-time point lacks its meaning. Such theories, when modelled with the help of commutative space-time endowed 
with star product, lead at once to nonlocal Lagrangians. If noncommutativity involves time variable the theory becomes nonlocal in time and is 
plagued with unitarity and causality problems, at least when quantized with the help of naive Feynman rules \cite{b7}. There exist alternative 
quantization schemes which seem to cure the untarity problems \cite{b8}; however, they are claimed to lead to new troubles \cite{b9}.

In view of this state of art it seems necessary to reconsider the quantization problem starting from first principles. First step is to 
put the theory in Hamiltonian form. The relevant framework is provided by Ostrogradski formalism \cite{b10} for higher derivative theories and 
its sophisticated version \cite{b11}  for nonlocal ones. The main problem with such procedures is that the resulting Hamiltonians are necessarily 
unbounded from below due to their behaviour at the infinity of phase space. This implies that the quantum theory, if exists, has no stable 
ground state.

Still, some hope exists because, in most interesting cases, the nonlocality enters only throught interaction term. Then one can pose the problem 
of quantizing the perturbative sector of the theory \cite{b4}, \cite{b12}. The initial value problem for perturbative solutions involves 
basic variables and their first time derivatives so the phase space for such solutions resembles the standard one. Moreover, for perturbation 
theory only the vicinity of the phase space is relevant and we can hope that Hamiltonian is here bounded from below leading to stable perturbative 
vacuum.

 In the present paper, inspired by Refs. \cite{b4} 
and \cite{b12}, we study in some detail the Hamiltonian formalism and quantization for simple system described by the Lagrangian containing second time 
derivative in the interaction term. In Sec.II we show in full generality that the perturbative sector of our theory can be described with the 
help of Dirac method. There are two constraints of second kind which allow to eliminate perturbatively Ostrogradski momenta in favour of coordinates 
$q$\ and $\dot q$. Dirac bracket $\{q,\dot q\}_D$\ can be then perturbatively computed to arbitrary order in coupling constant. It is, however, 
rather complicated. Therefore, in Sec.III we analize the possibility of simplifying the form of Dirac bracket and the Hamiltonian. We show 
that it is indeed possible to define perturbatively, order by order, the new variables $x,\dot x$\ such that: $(i)$\ the Dirac bracket 
takes standard form $\{x,\dot x\}=1, \;\; (ii) $\ the Hamiltonian is the sum of kinetic and potential energy. We show that there is a large freedom 
in defining $x$\ and $\dot x$\ obeying $(i),  (ii)$; in fact, at any order of perturbative expansion for $x$\ one can add many terms with 
new, also dimensionful, constans. These constants are spurious in the sense that they disappear after coming back to the original dynamical variables. 
However, this might be not the case in quantum theory as we explain in Sec.VI. Sec.IV is devoted to the special case of homogeneous (monomial) 
potentials. The form of the transformation $(q,\dot q)\rightarrow (x,\dot x)$\ is studied in some detail. In particular, it is shown that if the 
degree of homogenity is odd the above transformation can be chosen such that the resulting Hamiltonian is parity invariant. This implies that the 
initial theory, when restricted to the perturbative sector, posses some complicated discrete symmetry. The form of symmetry transformation can be 
determined, order by order; however, we would like to have simpler and more straightforward explanation of its emergence. In Sec. V we study the 
simplest  example of homogeneous potential of third degree, considered already in Refs. \cite{b4} and \cite{b12}. We find explicitly, up to 
fourth order, the transformation relating $q$\ and $x$\ as well as the Hamiltonian to this order, expressed in terms of $x,\dot x$\ variables. 
It appears that the resulting parity invariant potential is positive term by term, up to fourth order. On the other hand, the initial Hamiltonian, 
considered to the same order, is not positively definite. There is no contradiction here because our expansions are at best asymptotic and valid at 
vicinity of phase space. Moreover, we do not know whether the property of positivity of parity invariant potential persists in higher orders. If 
this the case, the theory is perturbatively stable.

Finally, in Sec.VI we study the quantum theory of the system described in Sec.V. To this end we consider (up to the second order) the transformation 
converting the Dirac bracket and the Hamiltonian into the standard form. As it has been already stressed such a transformation is not uniquely 
defined. We consider a one - parameter family of transformations and show that for different values of the parameter the resulting quantum 
theories are not equivalent. Specifically, the energy eigenvalues differ by an overall constant. Therefore, the additional parameter, spurious in the 
classical case, becomes meaningful when quantum corrections are taken into account. This shows that the quantization is subtle and ambiguous procedure.

\section{Hamiltonian formalism for the perturbative sector}
Let us consider the following Lagrangian
\begin{eqnarray}
L=\frac{1}{2}\dot q^2-\frac{\omega ^2}{2}q^2-gV(q,\dot q,\ddot q),\;\;\;\;\frac{\partial ^2V}{\partial \ddot q^2}\neq 0   \label{w1}
\end{eqnarray}
It depends on second derivative $\ddot q$; however, $\ddot q$\ enters $L$\ only throught $V$\ which, in turn, is multiplied by the 
coupling constant $g$. \\
The corresponding Euler-Lagrange equation reads
\begin{eqnarray}
\ddot q+\omega ^2q+g\left(\frac{\partial V}{\partial q}-\frac{d}{dt}\left(\frac{\partial V}{\partial \dot q}\right)+
\frac{d^2}{dt^2}\left(\frac{\partial V}{\partial \ddot q}\right)\right)=0     \label{w2}
\end{eqnarray}
This is a fourth-order differential equation. In order to obtain a unique solution one has to impose the initial conditions 
involving $q$\ and its first three derivatives. Correspondingly, the phase space of the system must be fourdimensional.

The canonical formalism for our system can be introduced according to the Ostrogradski prescription \cite{b10}. To this end we define the canonical 
variables
\begin{eqnarray}
&& q_1=q, \;\;\;\; q_2=\dot q     \label{w3}  \\
&& p_1=\frac{\delta L}{\delta \dot q}\equiv \frac{\partial L}{\partial \dot q}-\frac{d}{dt}\left(\frac{\partial L}{\partial \ddot q}\right)=
\dot q+g\left(-\frac{\partial V}{\partial \dot q}+\frac{d}{dt}\left(\frac{\partial V}{\partial \ddot q}\right)\right)\equiv 
p_1(q,\dot q,\ddot q,\stackrel{\ldots}{q})   \nonumber  \\
&& p_2=\frac{\delta L}{\delta \ddot q}\equiv \frac{\partial L}{\partial \ddot q}=-g\frac{\partial V}{\partial \ddot q}\equiv p_2(q,\dot q,\ddot q) \nonumber
\end{eqnarray}
and the Hamiltonian
\begin{eqnarray}
H\equiv p_1q_2+p_2\ddot q(q_1,q_2,p_2)-L(q_1,q_2,\ddot q(q_1,q_2,p_2))  \label{w4}
\end{eqnarray}
where $\ddot q(q_1,q_2,p_2)$\ is the solution to the last eq.(\ref{w3}).

The main disadvantage of $H$\ is that $p_1$\ enters it linearly so it is unbounded from below; the system is unstable. One can try to cure this 
by imposing constraints confining the system to some submanifold of phase space. A natural choice is to consider only perturbative solutions 
to eq.(\ref{w2}). Due to the fact that third and fourth derivatives enter only the terms multiplied by the coupling constant, the perturbative 
solution is uniquely determined by imposing the initial conditions on $q$\ and $\dot q$. In particular, higher derivatives can be expressed in 
terms of $q$\ and $\dot q$. In fact, one can write \cite{b12}
\begin{eqnarray}
&& \ddot q=f(q,\dot q)    \nonumber  \\
&& \stackrel{\ldots}{q}=\left(\dot q\frac{\partial }{\partial q}+f(q,\dot q)\frac{\partial }{\partial \dot q}\right)f\equiv Df \label{w5}  \\
&&  \;\;\;\; \vdots   \nonumber  \\
&& q^{(n)}=D^{n-2}f    \nonumber
\end{eqnarray}

The form of $f(q,\dot q)$\ is determined by demanding that it is consistent with Euler-Lagrange equations. Let $F(q,\dot q,\ddot q,...,q^{(n)})$\ 
be any function. Define \cite{b12}
\begin{eqnarray}
[F](q,\dot q)\equiv F(q,\dot q,Df,...,D^{n-2}f)   \label{w6}
\end{eqnarray}
Some properties of the bracket $[\;\cdot  \;]$\ are discussed in Appendix. \\
The consistency condition for $f$\ reads
\begin{eqnarray}
\left[\ddot q+\omega ^2q+g\left(\frac{\partial V}{\partial q}-\frac{d}{dt}\left(\frac{\partial V}{\partial \dot q}\right)+\frac{d^2}{dt^2}\left(
\frac{\partial V}{\partial \ddot q}\right)\right)\right]=0   \label{w7}
\end{eqnarray}

Assume now that we have found some $f(q,\dot q)$\ obeying (\ref{w7}). The definitions of $p_{1,2}$\ can be now converted into constraints
\begin{eqnarray}
&& \varphi _1\equiv p_1-[p_1(q,\dot q,\ddot q,\stackrel{\dots}{q})](q_1,q_2)=0   \label{w8} \\
&& \varphi _2\equiv p_2-[p_2(q,\dot q,\ddot q)](q_1,q_2)=0  \nonumber
\end{eqnarray}
Differentiating $\varphi _1,\varphi _2$\ with respect to time and using (\ref{w7}) and (\ref{w8}) we find that there are no secondary 
constraints.\\
The constraints $\varphi _1,\varphi _2$\ are second class ones:
\begin{eqnarray}
\{\varphi _1,\varphi _2\}=-g\frac{\partial }{\partial q_1}\left[\frac{\partial V}{\partial \ddot q}\right]-g\frac{\partial }{\partial q_2}
\left[-\frac{\partial V}{\partial \dot q}+\frac{d}{dt}\left(\frac{\partial V}{\partial \ddot q}\right)\right]  \label{w9}
\end{eqnarray}
Due to the form of constraints $\varphi _i$, the momenta $p_i$\ can be expressed in terms of $q_1$\ and $q_2$\ which parametrize the 
reduced phase space. Dirac bracket reads
\begin{eqnarray}
\{A,B\}_D=\{A,B\}+\{A,\varphi _1\}\{\varphi _1,\varphi _2\}^{-1}\{\varphi _2,B\}-\{A,\varphi _2\}\{\varphi _1,\varphi _2\}^{-1}\{\varphi _1,B\} \label{w10}
\end{eqnarray}
In particular
\begin{eqnarray}
\{q_1,q_2\}_D=-\{\varphi _1,\varphi _2\}^{-1}   \label{w11}
\end{eqnarray}
The same result is obtained by considering the symplectic form
\begin{eqnarray}
\Omega =dp_1\wedge dq_1+dp_2\wedge dq_2    \label{w12}
\end{eqnarray}
reduced to our submanifold. Indeed, we find
\begin{eqnarray}
\Omega _{red}=\left(\frac{\partial [p_1]}{\partial q_2}-\frac{\partial [p_2]}{\partial q_1}\right)dq_2\wedge dq_1  \label{w13}
\end{eqnarray}
so that
\begin{eqnarray}
\{q_1,q_2\}_{red}=\left(\frac{\partial [p_1]}{\partial q_2}-\frac{\partial [p_2]}{\partial q_1}\right)^{-1}  \label{w14}
\end{eqnarray}
which, by eq.(\ref{w9}), coincides with (\ref{w11}). One can also check the validity of Hamiltonian equations. It is convenient to come back to initial 
notation $q_1=q, \;\; q_2=\dot q$. Simple computation gives
\begin{eqnarray}
&& \frac{\partial [H]}{\partial q}=-f(q,\dot q)\{q,\dot q\}_D^{-1}   \label{w15} \\
&& \frac{\partial [H]}{\partial \dot q}=\dot q\{q,\dot q\}_D^{-1}     \nonumber
\end{eqnarray}
where eq.(\ref{w7}) has been used. Now, first Hamiltonian equation
\begin{eqnarray}
\dot q=\{q,[H]\}_D  \label{w16}
\end{eqnarray}
gives the identity $\dot q=\dot q$\ while the second one
\begin{eqnarray}
\ddot q=\{\dot q,[H]\}_D  \label{w17}
\end{eqnarray} 
leads to contraint equation
\begin{eqnarray}
\ddot q=f(q,\dot q)   \label{w18}
\end{eqnarray}
\section{Simplifying dynamics}
The form of reduced dynamics presented above is rather complicated; in particular, due to the nontrivial form of basic Poisson (Dirac) 
bracket (\ref{w11}) the quantization poses nontrivial ordering problem. In order to avoid this problem one can adopt the following strategy
 \cite{b4}, \cite{b12}: 
instead of direct quantization one first makes Darboux transformation which simplifies $\Omega _{red}, \;\Omega _{red}=d\dot x\wedge dx$. 
Such a transformation is not unique; in fact, it is defined up to a canonical transformation. The question arises whether this freedom can be 
used to simplify also the Hamiltonian or even to put it in standard form: kinetic plus potential energy.

In order to analyse this problem we start with the lowest order approximation. Let us first note that for the Lagrangian (\ref{w1}) the zeroth-order 
approximation to $f(q,\dot q)$\ reads
\begin{eqnarray}
f_0(q,\dot q)=-\omega ^2q    \label{w19}
\end{eqnarray}
The corresponding approximation to the time-derivative operator $D$\ will be denoted by $D_0$,
\begin{eqnarray}
D_0\equiv \dot q\frac{\partial }{\partial q}-\omega ^2q\frac{\partial }{\partial \dot q}    \label{w20}
\end{eqnarray}
Finally, $[\;\cdot \;]_0$\ denotes $[\;\cdot \;]$\ given by eq.(\ref{w6}) with $D$\ replaced by $D_0$. \\
Our aim is to define the transformation $(q,\dot q)\rightarrow (x,\dot x)$\ simplifying both Dirac bracket and Hamiltonian. To the first 
order in $g$\ one can write
\begin{eqnarray}
q=x+gm(x,\dot x)   \label{w21}
\end{eqnarray}
or
\begin{eqnarray}
x=q-gm(q,\dot q)   \label{w22}
\end{eqnarray}
To this order we have also
\begin{eqnarray}
\dot x=\dot q-gD_0m(q,\dot q)     \label{w23}
\end{eqnarray}
or
\begin{eqnarray}
\dot q=\dot x+gD_0m(x,\dot x)    \label{w24}
\end{eqnarray}
where $D_0$\ on the right-hand side of eq.(\ref{w24}) is given by eq.(\ref{w20}) with $q$\ replaced by $x$. \\
We start by writing the reduced symplectic form to the first order in $g$:
\begin{eqnarray}
\Omega _{red(1)}=\left(1+g\left(\frac{\partial }{\partial \dot q}\left(-\left[\frac{\partial V}{\partial \dot q}\right]_0+\left[\frac{d}{dt}
\left(\frac{\partial V}{\partial \ddot q}\right)\right]_0\right)+\frac{\partial }{\partial q}\left[\frac{\partial V}{\partial \ddot q}\right]_0
\right)\right)d\dot q\wedge dq    \label{w25}
\end{eqnarray}
We are looking for $m(x,\dot x)$\ such that the transformations (\ref{w21}), (\ref{w24}) lead to $\Omega _{red(1)}=d\dot x\wedge dx$. As a result of 
simple computation we obtain the following equation for $m(x,\dot x)$\
\begin{eqnarray}
\frac{\partial (D_0m)}{\partial \dot x}+\frac{\partial m}{\partial x}+\frac{\partial }{\partial \dot x}\left(-\left[\frac{\partial V}{\partial \dot x}
\right]_0+\left[\frac{d}{dt}\left(\frac{\partial V}{\partial \ddot x}\right)\right]_0\right)+\frac{\partial }{\partial x}\left(\left[\frac{\partial V}
{\partial \ddot x}\right]_0\right)=0    \label{w26}
\end{eqnarray}
which we rewrite as
\begin{eqnarray}
\frac{\partial }{\partial \dot x}\left(D_0\left(m+\left[\frac{\partial V}{\partial \ddot x}\right]_0\right)-\left[\frac{\partial V}
{\partial \dot x}\right]_0\right)+\frac{\partial }{\partial x}\left(m+\left[\frac{\partial V}{\partial \ddot x}\right]_0\right)=0    \label{w27}
\end{eqnarray}
Eq.(\ref{w27}) implies that
\begin{eqnarray}
&& m+\left[\frac{\partial V}{\partial \ddot x}\right]_0=\frac{\partial \Phi (x,\dot x)}{\partial \dot x}   \nonumber \\
&& D_0\left(m+\left[\frac{\partial V}{\partial \ddot x}\right]_0\right)-\left[\frac{\partial V}{\partial \dot x}\right]_0=
-\frac{\partial \Phi (x,\dot x)}{\partial x}    \label{w28}
\end{eqnarray}
for some function $\Phi $. By virtue of (\ref{w28}) $\Phi $\ obeys
\begin{eqnarray}
\frac{\partial (D_0\Phi )}{\partial \dot x}=\left[\frac{\partial V}{\partial \dot x}\right]_0=\frac{\partial [V]_0}{\partial \dot x}  \label{w29}
\end{eqnarray}
or
\begin{eqnarray}
D_0\Phi =[V]_0-\tilde V(x),   \label{w30}
\end{eqnarray}
$\tilde V(x)$\ being an arbitrary (up to now) function of $x$\ alone.

In order to answer the question whether we can always find, to the first order in $g$, the transformation which puts $\Omega _{red(1)}$\ in 
Darboux form let us note that we are looking for a transformation which, up to a given order, is defined globally in the phase space (optimally, 
$m(x,\dot x)$\ is some polynomial provided $V$\ is).\\
Let us introduce the polar coordinates
\begin{eqnarray}
&& x=r \cos \Theta   \label{w31} \\
&& \dot x=\omega r\sin \Theta   \nonumber
\end{eqnarray}
In terms of new coordinates eq.(\ref{w30}) reads
\begin{eqnarray}
\frac{\partial \Phi }{\partial \Theta }=-\omega ([V]_0(r \cos \Theta , \omega r \sin \Theta )-\tilde V(r \cos \Theta ))   \label{w32}
\end{eqnarray}
The right-hand side is some periodic function of $\Theta $. Therefore, one has
\begin{eqnarray}
\frac{\partial \Phi }{\partial \Theta }=\sum\limits_{n>0}(a_n(r)e^{in\Theta }+\overline{a_n(r)}e^{-in\Theta })+a_0(r)  \label{w33}
\end{eqnarray}
and $\Phi $\ is globally defined (periodic) provided $a_0(r)=0$. Consider the first term on the RHS of eq.(\ref{w32}). It is easy to see that 
the $\Theta $\ -independent term must be a function of $r^2$. Consider particular contribution of the form $\alpha _{k}r^{2k}$; it can be 
cancelled by the term $\alpha _{k}\frac{2^{2k}}{{2k\choose k}}x^{2k}$\ entering $\tilde V(x)$. We conclude that $\tilde V(x)$\ can be chosen in such a way 
that no $\Theta $\ -independent term appears on the RHS of eq.(\ref{w32}). With such a choice $m(x,\dot x)$, defined by first eq.(\ref{w28}), defines 
the transformation leading to standard symplectic form. Let us note that there is a considerable freedom in the choice of $\tilde V(x)$. 

In order to find the meaning of $\tilde V(x)$\ let us note that eqs.(\ref{w28}) and (\ref{w30}) imply the following identity
\begin{eqnarray}
(D_0^2+\omega ^2)m+\left[\frac{\partial V}{\partial x}\right]_0-\left[\frac{d}{dt}\left(\frac{\partial V}{\partial \dot x}\right)\right]_0+
\left[\frac{d^2}{dt^2}\left(\frac{\partial V}{\partial \ddot x}\right)\right]_0=\frac{\partial \tilde V}{\partial x}  \label{w34}
\end{eqnarray}
Now, by computing $\ddot x$\ from eq.(\ref{w23}), keeping terms up to the first order and using eq.(\ref{w33}) we arrive at the equation of 
motion for $x$:
\begin{eqnarray}
\ddot x=-\omega ^2x-g\frac{\partial \tilde V(x)}{\partial x}  \label{w35}
\end{eqnarray}
Therefore, due to $\{x,\dot x\}=1$, the Hamiltonian computed to the first order in $g$, has the form
\begin{eqnarray}
H=\left(\frac{1}{2}\dot x^2+\frac{\omega ^2x^2}{2}\right)+g\tilde V(x)  \label{w36}
\end{eqnarray}
Let us generalize our analysis to arbitrary order in $g$. To this end we write
\begin{eqnarray}
&& x=q-\sum\limits_{n=1}^\infty  g^nm_n(q,\dot q)\equiv q-M(q,\dot q)     \label{w37}  \\
&& \dot x=\dot q-\sum\limits_{n=1}^\infty  g^nDm_n(q,\dot q)\equiv \dot q-DM(q,\dot q)  \nonumber
\end{eqnarray}
Let us note that the second formula does not represent an explicit expansion in powers of coupling constant $g$. This is due to the fact that $D$\ 
itself contains $f(q,\dot q)$\ which is also given as power series in $g$.

Now, assuming that $\Omega _{red}$\ takes the standard form when expressed in terms of $x$\ and $\dot x$, we can write
\begin{eqnarray}
\Omega _{red}=d\dot x\wedge dx=\left(1-\frac{\partial DM}{\partial \dot q}-\frac{\partial M}{\partial q}+\frac{\partial DM}{\partial \dot q}
\frac{\partial M}{\partial q}-\frac{\partial DM}{\partial q}\frac{\partial M}{\partial \dot q}\right)d\dot q\wedge dq     \label{w38}
\end{eqnarray}
By virtue of eqs.(\ref{w3}), (\ref{w13}) and (\ref{w38}) we find that $M$\ obeys
\begin{eqnarray}
&& \frac{\partial DM}{\partial \dot q}+\frac{\partial M}{\partial q}-\left(\frac{\partial DM}{\partial \dot q}\frac{\partial M}
{\partial q}-\frac{\partial DM}{\partial q}\frac{\partial M}{\partial \dot q}\right)=  \nonumber \\
&& =g\left(\frac{\partial }{\partial \dot q}\left[\frac{\partial V}{\partial \dot q}-\frac{d}{dt}\left(\frac{\partial V}{\partial \ddot q}\right)
\right]-\frac{\partial }{\partial q}\left[\frac{\partial V}{\partial \ddot q}\right]\right)   \label{w39}
\end{eqnarray}
We want to solve eq.(\ref{w39}) perturbatively in $g$. Assume it holds up to n-th order and consider the $n+1$\ -st order. Note that the expression 
in the parenthesis is to be computed to n-th order only. Moreover, noting that $M$\ and $DM$\ are both at least $0(g)$\ we conclude that the equation 
for $n+1$\ -st order contribution to $M$\ reads
\begin{eqnarray}
\frac{\partial (D_0m_{n+1})}{\partial \dot q}+\frac{\partial m_{n+1}}{\partial q}=sum\; of\; known\; terms\;\equiv 
\frac{\partial ^2R_{n+1}}{\partial \dot q^2}   \label{w40}
\end{eqnarray}
where the known RHS we have rewritten for further convenience as a second derivative with respect to $\dot q$\ (which is always possible). \\
Eq.(\ref{w40}) can be written in the form
\begin{eqnarray}
\frac{\partial }{\partial \dot q}\left(D_0m_{n+1}-\frac{\partial R_{n+1}}{\partial \dot q}\right)+\frac{\partial m_{n+1}}{\partial q}=0   \label{w41}
\end{eqnarray}
Again we conclude that
\begin{eqnarray}
&& m_{n+1}=\frac{\partial \Phi _{n+1}}{\partial \dot q}   \label{w42} \\
&& D_0m_{n+1}-\frac{\partial R_{n+1}}{\partial \dot q}=-\frac{\partial \Phi _{n+1}}{\partial q}   \nonumber
\end{eqnarray}
for some $\Phi _{n+1}(q,\dot q)$. Eqs.(\ref{w42}) lead to the consistency condition for $\Phi _{n+1}$.
\begin{eqnarray}
D_0\frac{\partial \Phi _{n+1}}{\partial \dot q}+\frac{\partial \Phi _{n+1}}{\partial q}=\frac{\partial R_{n+1}}{\partial \dot q}   \label{w43}
\end{eqnarray}
or
\begin{eqnarray}
D_0\Phi _{n+1}(q,\dot q)=R_{n+1}(q,\dot q)+S_{n+1}(q)  \label{w44}
\end{eqnarray}
One can repeat the arguments used in the case of first order approximation. Namely, $\Phi _{n+1}$\ is globally well-defined provided $S_{n+1}$\ 
is chosen in such a way that no $\Theta $\ -independent term (cf. eqs.(\ref{w31})) appear on the RHS. This is always possible so we conclude 
that one can construct the standard canonical variables defined globally to arbitrary order in $g$.

Let us further note that the transformation $(q,\dot q)\rightarrow (x,\dot x)$\ of the phase space is defined in such a way that the second canonical 
variable continues to be the time derivative of the first one (for a given perturbative dynamics). Therefore, the first Hamilton equation 
is an identity which, due to $\{x,\dot x\}=1$, leads to the standard form of the Hamiltonian,
\begin{eqnarray}
H=\frac{1}{2}\dot x^2+\frac{1}{2}\omega ^2x^2+\tilde V(x;g)  \label{w45}
\end{eqnarray}
This can be also checked explicitly.

We have shown that, order by order, one can reduce to the standard form the perturbative sector of the dynamics defined by the Lagrangian (\ref{w1}).
\section{Homogeneous potentials}
Let us now consider the special case of homogeneous monomial potentials
\begin{eqnarray}
V(q,\dot q,\ddot q)=q^k\dot q^l\ddot q^m, \;\;\;\; m\geq 2;    \label{w46}
\end{eqnarray}
let us denote $a=k+l+m$. For dimensional reason one can write
\begin{eqnarray}
f(q,\dot q)=\sum\limits_{n=0}^\infty  g^nf_n(q,\dot q)    \label{w47}
\end{eqnarray}
where $f_n(q,\dot q)$\ are homogeneous polynomials of degree $n(a-2)+1$. \\
Also, one can write the perturbative expansions for other relevant quantities. First, we have
\begin{eqnarray}
\Omega _{red}=(1+\sum\limits_{n=1}^\infty  g^n\omega _n(q,\dot q))d\dot q\wedge dq   \label{w48}
\end{eqnarray}
where $\omega _n(q,\dot q)$\ are homogeneous polynomials of degree $n(a-2)$. On the other hand, we have seen in the last section that there is 
a large freedom in the choice of the functions $m_n(q,\dot q)$.
 Indeed, they are determined by the choice of $S_n(q)$\ (cf. eq.(\ref{w44})). There is only one condition restricting the admitted form of 
$S_n(q)$: the sum on the RHS should not contain the $\Theta $\ -independent term. This is rather weak condition which allows to add many terms 
(say, any homogeneous polynomial of odd degree) containing new (also dimensionful) parameters. However, one can show that it is always possible to choose 
the "minimal"$S_n's$\ in the sense that the only constants entering them are $g$\ and $\omega $. 
Assuming this is the case up the order $n$\ we conclude that $\frac{\partial ^2R_{n+1}}{\partial \dot q^2}$\ is homogeneous polynomial of degree 
$(n+1)(a-2)$\ depending only on one constant $\omega $. Therefore, $R_{n+1}$\ can be also chosen as homogeneous polynomial of degree 
$(n+1)(a-2)+2$\ containing only one dimensionful constant $\omega $. As a result, the $\Theta $\ -independent term in $R_{n+1}$\ must be of 
the form $r^{(n+1)(a-2)+2}$\ times a dimensionless constant. Then we can choose $S_{n+1}(q)$\ as proportional to $q^{(n+1)(a-2)+2}$\ and 
$\Phi _{n+1}$\ obeying eq.(\ref{w44}) can be taken as homogeneous polynomial of the same degree depending only on $\omega $. So, by first eq.(\ref{w42}) 
$m_{n+1}$\ is homogeneous of degree $n(a-2)+1$. This concludes the inductive proof.

With the minimal choice of the transformation (\ref{w37}) one can easily write out the general form of the potential $\tilde V(x;g)$\ entering 
the Hamiltonian (\ref{w45}); it reads
\begin{eqnarray}
\tilde V(x;g)=\sum\limits_{n=1}^\infty  v_ng^n\omega ^{(l+2m-2)n+2}x^{(a-2)n+2}     \label{w49}
\end{eqnarray}
Let us now consider the particular case of odd $a$. Notice that $R_n$\ is of degree $n(a-2)+2$\ which is odd for $n$\ odd. Therefore, 
$R_n$\ is then homogeneous polynomial of odd degree so it does not contain $\Theta $\ -independent term. So $S_n$\ can be chosen as 
$\alpha _nq^{n(a-2)+2}$\ with $\alpha _n$\ arbitrary (in particular, one can take $\alpha _n=0$\ ). It is not difficult to see that $\alpha _n$\ 
can be chosen perturbatively order by order so that the odd terms in the expansion (\ref{w49}) vanish. Indeed, let
\begin{eqnarray}
S_n=\alpha q^{n(a-2)+2}   \label{w50}
\end{eqnarray}
Once $S_n$\ is selected, one can define, via eqs.(\ref{w37}), (\ref{w42}) and (\ref{w44}), the variables $x_\alpha ,\dot x_\alpha $\ to n-th order. 
It is easy to see that the relation between $x_0,\dot x_0$\ (corresponding to the choice $\alpha =0$\ )   
and $x_\alpha ,\dot x_\alpha $, to the same order, reads
\begin{eqnarray}
&& x_0=x_\alpha +g^n\bigtriangleup m_\alpha (x_\alpha ,\dot x_\alpha )     \label{w51}  \\
&& \dot x_0=\dot x_\alpha +g^nD_0\bigtriangleup m_\alpha (x_\alpha ,\dot x_\alpha )  \nonumber
\end{eqnarray}
with
\begin{eqnarray}
&& \bigtriangleup m_\alpha =\frac{\partial \bigtriangleup \Phi _n}{\partial \dot x},\;\;\; D_0\bigtriangleup m_\alpha =\frac
{-\partial \bigtriangleup \Phi _n}{\partial x}    \nonumber \\
&& D_0\bigtriangleup \Phi _n=\alpha x^{n(a-2)+2}    \label{w52}
\end{eqnarray}
Therefore, adding the term (\ref{w50}) amounts to the following change of the Hamiltonian
\begin{eqnarray}
&& H=\frac{1}{2}\dot x_0^2+\frac{1}{2}\omega ^2x_0^2+\tilde V_n(x_0,g)\simeq \frac{1}{2}\dot x_\alpha ^2+\frac{1}{2}\omega ^2x_\alpha ^2+  \nonumber  \\
&& +\tilde V_n(x_\alpha ,g)+g^n\dot x_\alpha D_0\bigtriangleup m_\alpha +g^n\omega ^2x_\alpha \Delta m_\alpha = \nonumber  \\
&& =\frac{1}{2}\dot x_\alpha ^2+\frac{1}{2}\omega ^2x_\alpha ^2+\tilde V_n(x_\alpha ,g)-g^n\left(\dot x_\alpha \frac{\partial }
{\partial x_\alpha }-\omega ^2x_\alpha \frac{\partial }{\partial \dot x_\alpha }\right)\bigtriangleup \Phi _n=  \nonumber \\
&& =\left(\frac{1}{2}\dot x_\alpha ^2+\frac{1}{2}\omega ^2x_\alpha ^2+\tilde V_n(x_\alpha ,g)\right)-
\alpha g^nx_\alpha ^{n(a-2)+2}   \label{w53}
\end{eqnarray}
Adjusting properly $\alpha $\ one can cancel, order by order, all odd terms in $\tilde V(x,g)$.

Concluding, we find that for odd monomial $V(q,\dot q,\ddot q)$\ one can reduce, order by order, the perturbative potential $\tilde V(x;g)$\ to 
the form
\begin{eqnarray}
\tilde V(x;g)=\sum\limits_{k=1}^\infty v_{2k}g^{2k}\omega ^{2(l+2m-2)k+2}x^{2(a-2)k+2}   \label{w54}
\end{eqnarray}
Note that in this case the perturbative sector exhibits some discrete nonlinear symmetry. In fact, the resulting standard Hamiltonian is parity 
invariant: $x\rightarrow -x, \;\dot x\rightarrow -\dot x$\ is a symmetry. Then, expressed back in original variables, the parity transformation 
produces nonlinear symmetry defined order by order in coupling constant $g$.
\section{The simple example}
Let us consider a simple model studied already in Refs. \cite{b4}, \cite{b12}:
\begin{eqnarray}
L=\frac{1}{2}\dot q^2-\frac{1}{2}\omega ^2q^2-gq\ddot q^2    \label{w55}
\end{eqnarray}
It belongs to the class of models studied in the last section. Eq.(\ref{w55}) leads to the following equation of motion
\begin{eqnarray}
\ddot q+\omega ^2q+g(3\ddot q^2+4\dot q\stackrel{\ldots}{q}+2qq^{(IV)})=0  \label{w56}
\end{eqnarray}
The canonical variables read
\begin{eqnarray}
&& q_1=q,\;\;\; q_2=\dot q     \nonumber  \\
&& P_1=\dot q+2g(\dot q\ddot q+q\stackrel{\ldots}{q}) \label{w57}  \\
&& P_2=-2gq\ddot q  \nonumber
\end{eqnarray}
It is also straightforward to write out the 
Hamiltonian
\begin{eqnarray}
H=P_1q_2-\frac{P_2^2}{4gq_1}-\frac{1}{2}q_2^2+\frac{1}{2}\omega ^2q_1^2    \label{w58}
\end{eqnarray}
In order to perform the reduction to the perturbative sector we impose the constraint
\begin{eqnarray}
\ddot q=f(q,\dot q)    \label{w59}
\end{eqnarray}
Then, by virtue of eq.(\ref{w56}), $f(q,\dot q)$\ obeys
\begin{eqnarray}
&& f+\omega ^2q+g\left(3f^2+4\dot q\left(\dot q\frac{\partial f}{\partial q}+f\frac{\partial f}{\partial \dot q}\right)\right)+  \label{w60}  \\
&& +2q\left(\dot q^2\frac{\partial ^2f}{\partial q^2}+\dot q\frac{\partial f}{\partial q}\frac{\partial f}{\partial \dot q}+
f\frac{\partial f}{\partial q}+2\dot qf\frac{\partial ^2f}{\partial q\partial \dot q}+f\left(\frac{\partial f}{\partial \dot q}\right)^2+
 f^2\frac{\partial ^2f}{\partial \dot q^2}\right)=0     \nonumber
\end{eqnarray}
This equation, although quite complicated, can be solved perturbatively order by order in $g$. For example, to the third order in $g$\ one finds 
\begin{eqnarray}
&&f=-\omega ^2q-g(5\omega ^4q^2-4\omega ^2\dot q^2)+g^2(-76\omega ^6q^3+140\omega ^4q\dot q^2)+  \label{w61} \\
&&-g^3(1959\omega ^8q^4-6800\omega ^6q^2\dot q^2+736\omega ^4\dot q^4)  \nonumber
\end{eqnarray}
The constraints (\ref{w8}) take the form
\begin{eqnarray}
&& P_1-\dot q-2g\left(\dot q \dot f+q\dot q\frac{\partial f}{\partial q}+qf\frac{\partial f}{\partial \dot q}\right)\approx 0    \nonumber  \\
&& P_2+2gqf\simeq 0  \label{w62}
\end{eqnarray}
while $\Omega _{red}$\ is given by 
\begin{eqnarray}
&& \Omega _{red}= \label{w63}  \\
&& =\left(1+4gf+4gq\frac{\partial f}{\partial q}+2g\dot q\frac{\partial f}{\partial \dot q}+2gq\dot q\frac{\partial ^2f}
{\partial q\partial \dot q}+2gq\left(\frac{\partial f}{\partial \dot q}\right)^2+2gqf\frac{\partial ^2f}{\partial \dot q^2}\right)
d\dot q\wedge dq   \nonumber
\end{eqnarray}
Finally, the reduced Hamiltonian reads
\begin{eqnarray}
[H]=\frac{1}{2}\dot q^2+\frac{1}{2}\omega ^2q^2+g\left(-qf^2+2\dot q^2f+2q\dot q^2\frac{\partial f}{\partial q}+2q\dot qf\frac{\partial f}
{\partial \dot q}\right)  \label{w64}
\end{eqnarray}
Now, one can try to find perturbatively the "normal" coordinates $x,\dot x$. Following the method outlined in previous sections we found 
that, to the fourth order,
\begin{eqnarray}
&& q=x+g(\omega ^2x^2-2\dot x^2)+g^2\left(\frac{50}{3}\omega ^4x^3-18\omega ^2x\dot x^2\right)+ \nonumber  \\
&& +g^3\left(\frac{760}{3}\omega ^6x^4-716\omega ^4x^2\dot x^2-84\omega ^2\dot x^4\right)+   \label{w65} \\
&&+g^4\left(\frac{111422}{15}\omega ^8x^5-25928\omega ^6x^3\dot x^2+3030\omega ^4x\dot x^4\right)+0(g^5)  \nonumber
\end{eqnarray}
and
\begin{eqnarray}
[H]=\frac{1}{2}\dot x^2+\frac{1}{2}\omega ^2x^2+\frac{25}{6}g^2\omega ^6x^4+\frac{30136}{45}g^4\omega ^{10}x^6+0(g^6)   \label{w66}
\end{eqnarray}
We see that our perturbative Hamiltonian, when put in normal form, becomes positively defined, at least up to fourth order in $g$. 
We don't know whether this property persists in higher orders. Let us note that our reduced Hamiltonian (\ref{w64}) is not positive. For example, 
to the first order in $g$\ one finds from (\ref{w61}) and (\ref{w64})
\begin{eqnarray}
[H]=\frac{1}{2}\dot q^2+\frac{1}{2}\omega ^2q^2-g\omega ^2(\omega ^2q^3+4q\dot q^2)   \label{w67}
\end{eqnarray}
which is negative for large $q,\dot q$. \\
On the other hand, to the same order $[H]$, when expressed in terms of new coordinates, is simply the energy of harmonic oscillator. We conclude that, 
at best, we can expect that our series defining new coordinates are asymptotic (note that $[H]$, as given by eq.(\ref{w67}), becomes negative 
for $q,\dot q$\ of order $\frac{1}{g}$\ ).
\section{Quantum theory}
Our ultimate goal is to quantize the higher derivative dynamical system. The main disadvantage of the Hamiltonian formalism introduced by 
Ostrogradski is that some momenta enter the Hamiltonian linearly. Therefore, it is unbounded from below. Contrary to the case where the 
Hamiltonian is unbounded in small regions of phase space, this kind of unboundness cannot be cured with the help of uncertainty principle. 
As a result, no stable ground state can exist.

However, one can ask whether it is possible to quantize consistently the higher-derivative theory in the perturbative sector. The first 
trouble is related here with the complicated form of reduced symplectic structure. It is by far not sure whether one can find the proper 
ordering procedure which allows to convert complicated Poisson brackets into commutators obeying Jacobi identity.

The simplest way to define the perturbative quantum theory seems to be the following. First, we construct on the classical level the 
transformation in reduced phase space leading to the standard from of the Poisson bracket and the Hamiltonian. Then the quantization can be performed 
in a straightforward way. Moreover, if the classical Hamiltonian appears to be bounded from below, the quantum theory possess perturbatively stable 
ground state. Once the theory is quantized in "standard" coordinates one defines the quantum counterparts of initial variables by inverting 
(perturbatively) the classical map and choosing a definite ordering (for example, the Weyl one).

The main problem here is that such a procedure is by far not unique. In fact, we have seen in previous section that there is a large freedom in defining 
the classical transformation to standard coordinates. One can hardly believe that the quantum theories resulting from different choices of such 
transformations are equivalent. Moreover, in the process of defining the perturbative transformation from $q$\ - to $x$\ -variables one can introduce 
new (also dimensionful) constants. On the classical level they are spurious and disappear after coming back to original dynamical variables. 
This may be not the case after quantization has been performed and the additional parameters may appear to be relevant.

In order to illustrate this phenomenon let us go back to our simple model. Consider the transformation
\begin{eqnarray}
&& x=q+g(\beta \omega ^2q^2+(2\beta +4)\dot q^2)+  \nonumber \\
&&g^2((-2\beta -\frac{50}{3})\omega ^4q^3+(-32-2\beta ^2-24\beta )\omega ^2q\dot q^2)  \nonumber \\
&& \dot x=\dot q-2g(\beta +4)\omega ^2q\dot q+  
 g^2((4\beta ^2+22\beta -26)\omega ^4\dot qq^2-2(\beta ^2+4\beta )\omega ^2\dot q^3   \label{w68}
\end{eqnarray}
depending on one real parameter $\beta $. In terms of new variables the Hamiltonian takes the form
\begin{eqnarray}
[H]=\frac{1}{2}\dot x^2+\frac{1}{2}\omega ^2x^2-g(\beta +1)\omega ^4x^3+5g^2\left(\frac{1}{2}\beta ^2+\beta +\frac{4}{3}\right)\omega ^6x^4
+O(g^3)   \label{69}
\end{eqnarray}
For $\beta =-1$\ we obtain the parity invariant form.\\
Let us now compute the energies to the second order in $g$. Standard perturbation theory gives
\begin{eqnarray}
E_n=\hbar \omega \left(n+\frac{1}{2}\right)+\frac{25}{8}g^2\hbar ^2\omega ^4(n^2+(n+1)^2)+\frac{1}{2}g^2\hbar ^2\omega ^4(\beta +1)^2   \label{w70}
\end{eqnarray}
We see that the energy eigenvalues depend on $\beta $, although it is only an overall shift. It is interesting to note that the energies take minimal 
values in the parity - invariant case.

The ambiguity considered above is rather mild. We could add other terms, much more complicated and containing new dimensionful constants. 
Let us remind that the only condition imposed, order by order, on new $S_n$\ (cf. eq.(\ref{w44})) is that the RHS contain no $\Theta $\ -independent 
terms. Keeping this in mind one can easily understand that the resulting form of standard Hamiltonian can vary considerably depending on the 
particular transformation chosen. This may have strong impact on the form of energy spectrum. The resulting quantum theories become nonequivalent. 
This effect can be ultimately ascribed to the ordering problem.
\section{Appendix}
The main property of the symbol $[\;\cdot \;]$\ introduced in Sec. II is expressed by the equation \cite{b12}
\begin{eqnarray}
\left[\frac{d[F]}{dt}\right]=D[F]=\left[\frac{dF}{dt}\right] \label{w71}
\end{eqnarray}
To see this let us note that \cite{b12}
\begin{eqnarray}
\frac{d[F]}{dt}=\left[\frac{dF}{dt}\right]+\frac{\partial [F]}{\partial \dot q}(\ddot q-f)     \nonumber
\end{eqnarray}Due to $[\ddot q-f]=0$\ we find
\begin{eqnarray}
\left[\frac{d[F]}{dt}\right]=\left[\frac{dF}{dt}\right]   \nonumber
\end{eqnarray}
Also $D[F]=[D[F]]$\ and $D[F]=[D[F]]=\left[\frac{d[F]}{dt}\right]=\left[\frac{dF}{dt}\right] $\  \\
Iterating (\ref{w71}) one obtains
\begin{eqnarray}
\left[\frac{d^2F}{dt^2}\right]=\left[\frac{d}{dt}\left[\frac{dF}{dt}\right]\right]= \left[\frac{d}{dt}\left[\frac{d[F]}{dt}\right]
\right]=\left[\frac{d^2[F]}{dt^2}\right]   \nonumber
\end{eqnarray}
and
\begin{eqnarray}
D^2[F]=D[D[F]]=\left[\frac{d}{dt}(D[F])\right]=\left[\frac{d^2[F]}{dt^2}\right]   \nonumber
\end{eqnarray}
Therefore
\begin{eqnarray}
\left[\frac{d^n[F]}{dt^n}\right]=D^n[F]=\left[\frac{d^nF}{dt^n}\right]  \nonumber
\end{eqnarray}

\end{document}